\documentclass[conference]{IEEEtran}
\IEEEoverridecommandlockouts
\usepackage{cite,citesort}
\usepackage{amsmath,amssymb,amsfonts}
\usepackage{algorithmic}
\usepackage{graphicx}
\usepackage{textcomp}
\usepackage{xcolor}
\usepackage{url}
\def\BibTeX{{\rm B\kern-.05em{\sc i\kern-.025em b}\kern-.08em
    T\kern-.1667em\lower.7ex\hbox{E}\kern-.125emX}}

\begin{document}

\title{Affine Subspace Models and Clustering\\ for Patch-Based Image Denoising
\thanks{Portions of T. Wickremasinghe's work were completed while at University of Moratuwa, Sri Lanka.}
}

\author{\IEEEauthorblockN{Tharindu Wickremasinghe}
\IEEEauthorblockA{\textit{Electrical and Computer Engineering} \\
\textit{Purdue University}\\
West Lafayette, IN \\
lwickrem@purdue.edu}
\and
\IEEEauthorblockN{Marco F. Duarte}
\IEEEauthorblockA{\textit{Electrical and Computer Engineering} \\
\textit{University of Massachusetts}\\
Amherst, MA \\
mduarte@ecs.umass.edu}
}

\maketitle

\begin{abstract}
Image tile-based approaches are popular in many image processing applications such as denoising (e.g., non-local means). A key step in their use is grouping the images into clusters, which usually proceeds iteratively splitting the images into clusters and fitting a model for the images in each cluster. Linear subspaces have emerged as a suitable model for tile clusters; however, they are not well matched to images patches given that images are non-negative and thus not distributed around the origin in the tile vector space. We study the use of affine subspace models for the clusters to better match the geometric structure of the image tile vector space. We also present a simple denoising algorithm that relies on the affine subspace clustering model using least squares projection. We review several algorithmic approaches to solve the affine subspace clustering problem and show experimental results that highlight the performance improvements in clustering and denoising.
Code is available at \url{https://github.com/Tharindu-Nirmal/psp-affine-clustering}.
\end{abstract}


\section{Introduction}
Image denoising has recently re-emerged as a topic of interest due to its versatility and applicability as a tool towards a variety of image processing problems such as restoration, super-resolution, and compressive imaging~\cite{romano2017little,reehorst2018regularization,venkatakrishnan2013plug,milanfar2012tour}. In particular, so-called {\em non-local} models deviate from legacy approaches that consider each image pixel's local neighborhood to obtain estimates of the true pixel value; instead, they focus on the similarity between image patches (also often referred to as tiles) to drive the denoising process via weighted averaging~\cite{buades2005review,dabov20073d,mairal2009non,lebrun2013nonlocal,wu2013probabilistic}. A key step in such non-local methods is to perform a clustering of the image patches in order to determine which patches should be involved in each weighted averaging. While baseline methods such as $k$-means clustering require careful tuning of their parameters and have performance highly dependent on their initialization, the large scale of image tile datasets makes approaches based on subspace models attractive~\cite{zhang2010two,vidal2011subspace,deledalle2011image,dabov2009bm3d,bradley2000k}. Subspace models also underline a variety of contemporary approaches for image modeling and compression based on sparsity~\cite{wang2022convergence,chen2015external}. A variety of approaches to subspace clustering have been used in the literature, many of them based on optimization techniques~\cite{soltanolkotabi2014l1,elhamifar2013sparse}.

However, the use of linear subspaces is best chosen when the elements of each cluster are zero-mean, as all subspaces include the origin. In practice, it is more accurate to describe patches as being contained in {\em affine} subspaces, where the shift from a linear subspace allows for a non-zero average of the patches in a cluster. 
While it is common wisdom that simply adding a new dimension to each linear subspace (to account for the presence of the nonzero mean) alleviates the need for affine subspace modeling~\cite{vidal2011subspace}, there are some indications in the literature that explicitly enforcing affineness can increase the power of the subspace model~\cite{elhamifar2013sparse,ji2014efficient,li2018on,you2019affine}. Thus, in this paper we describe approaches that enforce the affineness constraint in subspace clustering for image patches, and subsequently capitalize on the potential improvements in clustering performance down the image processing pipeline --- in particular, by employing affine subspace clustering on the patches of a noisy image, leveraging the affine subspace model for each image tile within an image denoising algorithm that achieves improved performance, and by applying such improved denoisers in other image processing applications.


\section{Background}
\label{sec:bg}

\subsection{Notation}
We denote an $N$-pixel image of interest as $\mathbf{x}$, which may denote its representation either as a vector in $\mathbb{R}^N$ or as a matrix in $\mathbb{R}^{W\times H}$, where $W\cdot H = N$ are the image width and height, respectively. Its pixels are indexed by either a scalar in the range $1,\ldots,N$ or an ordered pair $(w,h)$ denoting the pixel coordinates; we denote both of these options for indexing by a single letter in the sequel for brevity. The image is polluted by additive noise $\mathbf{y}=\mathbf{x}+\mathbf{n}$. We denote a patch of the image $\mathbf{x}$ centered around pixel coordinate $i$ and of size $L\times L$ as $\bar{\mathbf{x}}[i]$.

\subsection{Nonlocal methods for image denoising}
\label{sec:nlmeans}
In non-local means denoising~\cite{buades2005review,dabov20073d,mairal2009non,chatterjee2011patch}, the denoised image $\hat{\mathbf{x}}$ is defined to have patches $\bar{\hat{\mathbf{x}}}[i] = \sum_{j \in \mathcal{N}_i}w(i,j)\bar{\mathbf{y}}[j]$ (with suitable averaging of overlapping patches if needed), where the family of weights $\{w(i,j)\}_{j\in \mathcal{N}_i}$ depends on the similarity between the image patches centered at locations $i$ and $j$, and $\mathcal{N}_i$ denotes the coordinates for the image patches that are judged similar to the patch at coordinate $i$ (e.g., the patch's neighborhood). The success of the algorithm hinges on appropriate selections of the weights $w$ and the neighborhoods $\mathcal{N}_i$, with common selections relying on neighborliness in patch space or clustering approaches~\cite{mairal2009non,chen2015external,chan2014monte}. Unfortunately, these clustering algorithms are highly sensitive to their randomized initializations or choice of initial parameters.

\subsection{Geometric models for image patches}
In the past, subspace models have been popularized for clustering, including for image tiles~\cite{elad2006image,vidal2011subspace}. Subspace clustering, which fits a subspace for each cluster obtained, provides an attractive option for partitioning large-scale datasets. Subspace models also underlie approaches based on sparsity-inducing transforms or representations, given that such transforms and representations provide a model for the data of interest as a union of subspaces~\cite{portilla2003image,elhamifar2011sparsity,deledalle2018image,liu2010robust,roth2005fields,van2014student}.

More recently, several approaches efficiently implement subspace clustering as an optimization problem by leveraging spectral clustering of an affinity score matrix~\cite{elhamifar2013sparse,gu2014weighted,soltanolkotabi2014l1,wang2022convergence,li2015structured}. 
{\em Self-representation} affinity (also called self-expressiveness) uses the fact that when the patches form a union of (linear or affine) subspaces each patch $\bar{\mathbf{x}}[i]$ can be represented as a linear combination of other patches contained in the same subspace. The linear combination representation coefficients provide the similarity weights. 
It is then possible to separate the patches into the different subspaces efficiently by applying spectral clustering on the affinity score matrix~\cite{zoran2011learning,elhamifar2013sparse}. 

\subsection{Affine subspaces for image patches}
When considering subspaces as models for image patches, it is worth noting that images are non-negative, and the mean of the data contained in each subspace will be nonzero. Thus, linear subspace models may not be best suited for their representation due to need for any linear subspace to contain the origin of the vector space. In contrast, affine subspaces can more efficiently account for the structure introduced by non-negativity, and so we consider their use in the various applications of image processing where linear subspace models are applied to image patches. 

Although it is possible to expand these linear subspace models by increasing the dimension of each subspace to account for the nonzero mean of the data, a subspace model still assumes that the coefficient for the mean is arbitrary; furthermore, the mean itself has to be learned or fit separately. Previous work has shown that the performance of this extended linear model can be lower than that of affine models, in particular when the data dimensionality is not high~\cite{elhamifar2013sparse,ji2014efficient,li2018on,you2019affine}. We therefore consider the use of affine subspaces in the clustering approaches applied to image patches, each of which contains only a few pixels.

\subsection{Optimization-based approaches for self-representation}
\label{sec:selfr}

Several approaches to obtain self-representations for sets of patches have been considered in the literature. More specifically, we consider the representation of patch $\bar{\mathbf{x}}$ (where we drop the coordinate for brevity) as a linear combination of the remaining patches collected as the columns of an {\em image patch matrix} $\mathbf{X}$, i.e., $\bar{\mathbf{x}}=\mathbf{Xc}$; once again, our expectation is that the patches $\bar{\mathbf{x}}$ in a given subspace have similar representations $\mathbf{c}$. The representations for the different patches are then collected as the columns of a self-representation matrix $\mathbf{C}$, which then provides an affinity matrix $\mathbf{W}=|\mathbf{C}|+|\mathbf{C}^T|$ to which spectral clustering is applied. In a nutshell, if one patch is present in the decomposition of another with a large coefficient magnitude, then the two patches will be considered to have high affinity.

The most common approach to obtain the representations $\mathbf{c}$ is known as basis pursuit denoising (BPDN), which promotes sparsity in the vector $\mathbf{c}$:
\begin{align}
\hat{\mathbf{c}}=\arg \min_\mathbf{c} \|\mathbf{c}\|_1~\mathrm{s.t.}~\|\mathbf{Xc}-\bar{\mathbf{x}}\|_2\le \sigma.
\label{eq:bpdn}
\end{align}
This approach is successful when the dimension of the subspaces is much smaller than the number of patches being considered (e.g., the number of columns in the matrix $\mathbf{C}$). However, BPDN requires an estimate of the goodness of fit of the self-representation model (e.g., a setting of the bound parameter $\sigma$).

\section{Self-Representations with\\ Affine Subspaces}
\label{sec:affine}

The self-representations obtained from (\ref{eq:bpdn}) can be ensured to model a patch decomposition captured by an affine subspace by requiring that the coefficients in the decomposition add up to one, i.e., we can add the constraint $\mathbf{1}^T\mathbf{c} = 1$ to the optimization problem above.
When $\mathbf{c}$ is non-negative ($\mathbf{c} \succcurlyeq 0$), this constraint is equivalent to $\|\mathbf{c}\|_1=\mathbf{1}^T\mathbf{c}$.
An efficient approach to enforcing this constraint is to use the non-negative lasso:
\begin{align}
\hat{\bf{c}}=\arg \min_\mathbf{c} \|\mathbf{Xc}-\bar{\mathbf{x}}\|_2~\mathrm{s.t.}~\|\mathbf{c}\|_1 \le \tau, \mathbf{c} \succcurlyeq 0.
\label{eq:nnclasso}
\end{align}
The non-negative lasso is a special instance of the so-called constrained lasso~\cite{gaines2018algorithms}. If the patch $\bar{\mathbf{x}}$ is indeed contained in an affine subspace, we can obtain the solution for the affine representation (e.g., one that obeys $\|\hat{\mathbf{c}}\|=\mathbf{1}^T\hat{\mathbf{c}}=1$) by setting $\tau = 1$. In this case, and in contrast to BPDN, the choice of parameter of the inequality constraint serves an additional purpose: establishing the affinity constraint for the desired solution.

The non-negative lasso is usually implemented as its Lagrangian relaxation 
\begin{align}
\hat{\mathbf{c}}=\arg \min_\mathbf{c} \frac{1}{2}\|\mathbf{Xc}-\bar{\mathbf{x}}\|_2^2+\alpha\|\mathbf{c}\|_1~\mathrm{s.t.}~ \mathbf{c} \succcurlyeq 0.
\label{eq:nnlasso}
\end{align}
For this formulation, it is necessary to fine-tune the weight $\alpha$ in order to meet the constraint $\|\mathbf{c}\|_1=\mathbf{1}^T\mathbf{c} = 1$; in some cases, we can only meet it approximately. As before, the affine representation vectors $\hat{\mathbf{c}}$ are collected as columns of an affinity matrix $\mathbf{C}$ to which spectral clustering is applied; the result provides a clustering of the patches themselves that leverages the affine subspace model.

Although both optimizations (\ref{eq:nnclasso}) and (\ref{eq:nnlasso}) are referred to in the literature as the non-negative lasso, we distinguish them by referring to (\ref{eq:nnclasso}) as the non-negative constrained (NNC) lasso and to (\ref{eq:nnlasso}) as the non-negative (NN) lasso, respectively.

\section{Image Denoising with\\ Subspace Patch Models}
\label{sec:ssden}

Under a union of subspaces model for the image patches, it is possible to develop a simple image denoising algorithm that assumes the image patches lie in a union of subspaces or, more specifically, that each cluster corresponds to a subspace. Under an additive white Gaussian noise (AWGN) model, the maximum likelihood estimate of an image patch corresponds to the least squares projection of the noisy image patch into the closest subspace (e.g, the one corresponding to its cluster). 

The proposed {\em patch subspace projection (PSP)} denoising procedure proceeds as follows. We assume a tile clustering has been obtained following one of the approaches of Section~\ref{sec:selfr}, and we learn the optimal subspace to model the images in each cluster; its dimensionality can be made data-dependent or be fixed {\em a priori}. First, we identify the subspace that is closest to the noisy image patch $\mathbf{y}$ (e.g., the subspace containing the point closest to the noisy image patch). Second, we compute the average of the patches in the subspace, denoted $\bar{\mathbf{x}}$, and subtract it from the noisy patch $\mathbf{y}_r =\mathbf{y}-\bar{\mathbf{x}}$. Third, we perform a principal component analysis projection of $\mathbf{y}_r$ into the mean-removed subspace to obtain the projection $\mathbf{x}_r$. Finally, we add the mean back to obtain the denoised patch $\hat{\mathbf{x}} = \mathbf{x}_r+\bar{\mathbf{x}}$. If the patches obtained from the image overlap, they can be merged via weighted averaging in a manner similar to what is done in non-local means~\cite{buades2005review}.

\section{Numerical Evaluation}
\label{sec:numerical}

To test the benefit of affineness in subspace models for patches, we compare the performance of nonlocal means denoisers (cf.\ Section~\ref{sec:nlmeans}) with the subspace-based denoising approach PSP (cf.\ Section~\ref{sec:ssden}) where the affinity matrix used in subspace clustering is computed using the three self-representation approaches described in Sections~\ref{sec:selfr} and~\ref{sec:affine}: BPDN, NNC lasso, and NN lasso. Note that BPDN does not enforce affineness, NNC lasso strictly enforces affineness, and NN lasso approximately enforces affineness. For these experiments, noisy input images $\mathbf{y}$ with AWGN at various levels are generated. We use ten Matlab demo images~\cite{matlabIPT_demoimages} 
which have size $256\times 256$ and pixel values in the range [0,255]. 
The images are tiled into square patches of size $8\times 8$ pixels each. Thereby, the noisy image patch matrix $\mathbf{Y}$ is of size $N\times d$, where $N=1024$ is the number of patches, each containing $d=64$ pixels.
All images are converted to 8-bit grayscale and degraded by adding independent zero-mean Gaussian noise with standard deviation $\sigma$ (on the 0–255 intensity scale), followed by clipping to [0,255]. For each $\sigma$, a fixed random seed is used to ensure reproducible noise realizations.

\begin{figure}[t!]
    \centering
    \includegraphics[width= 0.47\textwidth]{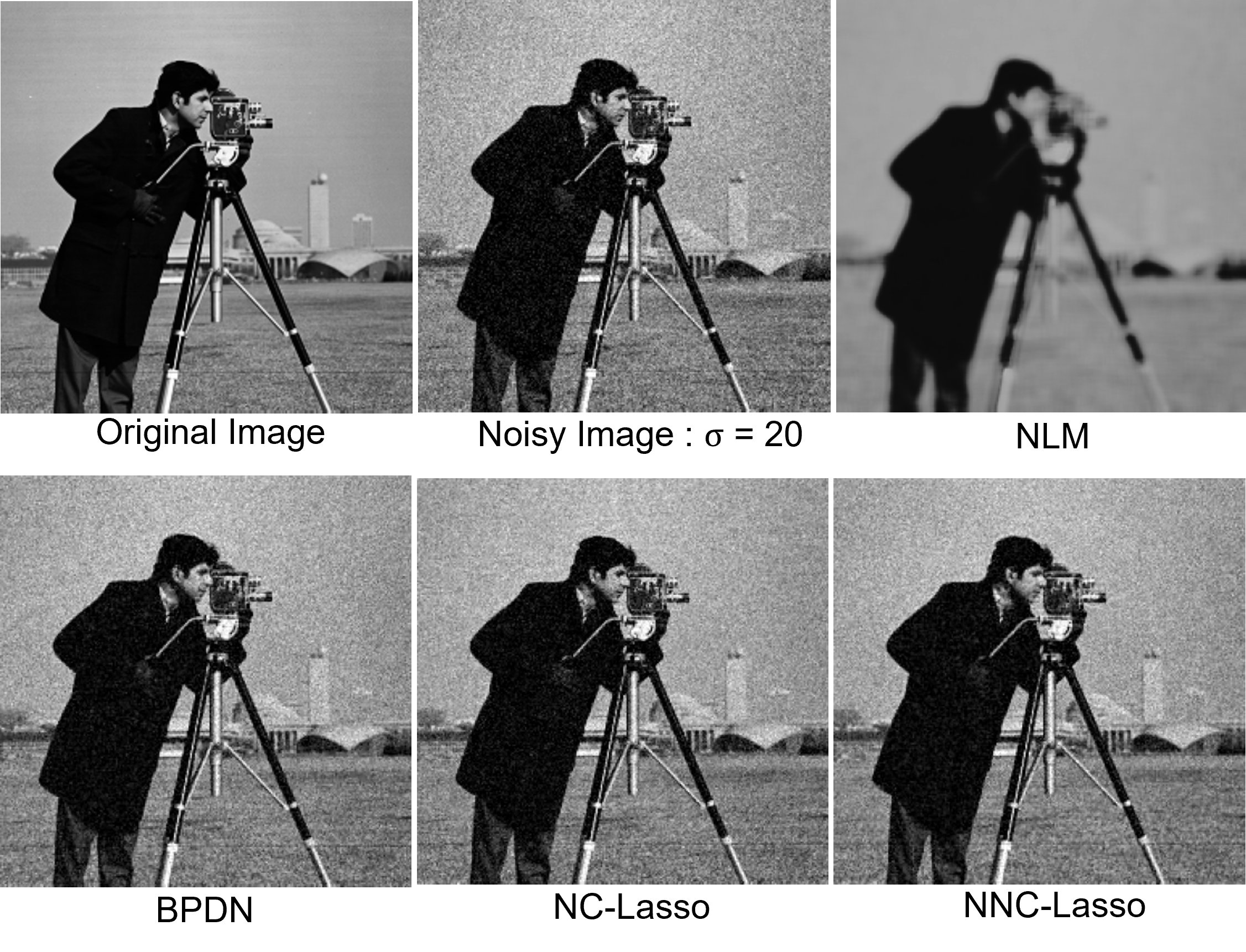}  
    \caption{Denoising example results for {\em Cameraman} image: original, noisy, and denoised images for each of the methods: BPDN, NC-Lasso, and NNC-Lasso. The noisy image has additive gaussian noise with $\sigma = 20$ pixels. Non-Local-Means (NLM) uses an estimate of the noise $\hat{\sigma}$ for the filter parameter.}
    \label{fig:cameraman}
\end{figure}

To replicate a practical scenario, we use the library \texttt{estimate\_sigma} from {\em scikit-image}~\cite{scikit-learn} to estimate the noise variance $\sigma$ from the noisy image $\mathbf{y}$. 
For the NNC lasso, we set $\tau = 1$ regardless of the input image as prescribed for affineness in Section~\ref{sec:affine}. 
The similarity matrices $\mathbf{W}_\mathrm{BPDN}, \mathbf{W}_\mathrm{NNC}$ obtained via BPDN and NNC lasso are both computed using solvers from the SPGL toolbox~\cite{Berg2011SparseOW}. 
For the NN lasso algorithm, we estimate the parameter $\alpha$ using a grid search over the set $[0.001, 0.01, 0.1, 1, 10]$; for each patch, we select the parameter that yields the sum of self-similarity coefficients closest to one. 
Experimentally, we observed that the choice of $\alpha$ does not affect this constraint beyond $\alpha = 10$, and thus it was selected as the largest parameter value of the search. The \texttt{ElasticNetCV} solver from {\em scikit-learn}~\cite{scikit-learn} was used to estimate $\mathbf{W}_\mathrm{NN}$ by setting the weight of the $\ell_2$-norm term to zero. For all methods,
 $\mathbf{W}$ is used as the pre-computed similarity matrix in {\em scikit-learn}'s \texttt{SpectralClustering} module~\cite{scikit-learn}. The choice of the number of clusters $K$ is data dependent; for our choice of images, we use $K$-means clustering and plot an elbow-curve to estimate a suitable number of clusters $\hat{K}=20$. 

\begin{figure}[t!]
    \centering
    
        \includegraphics[width=0.48\textwidth]{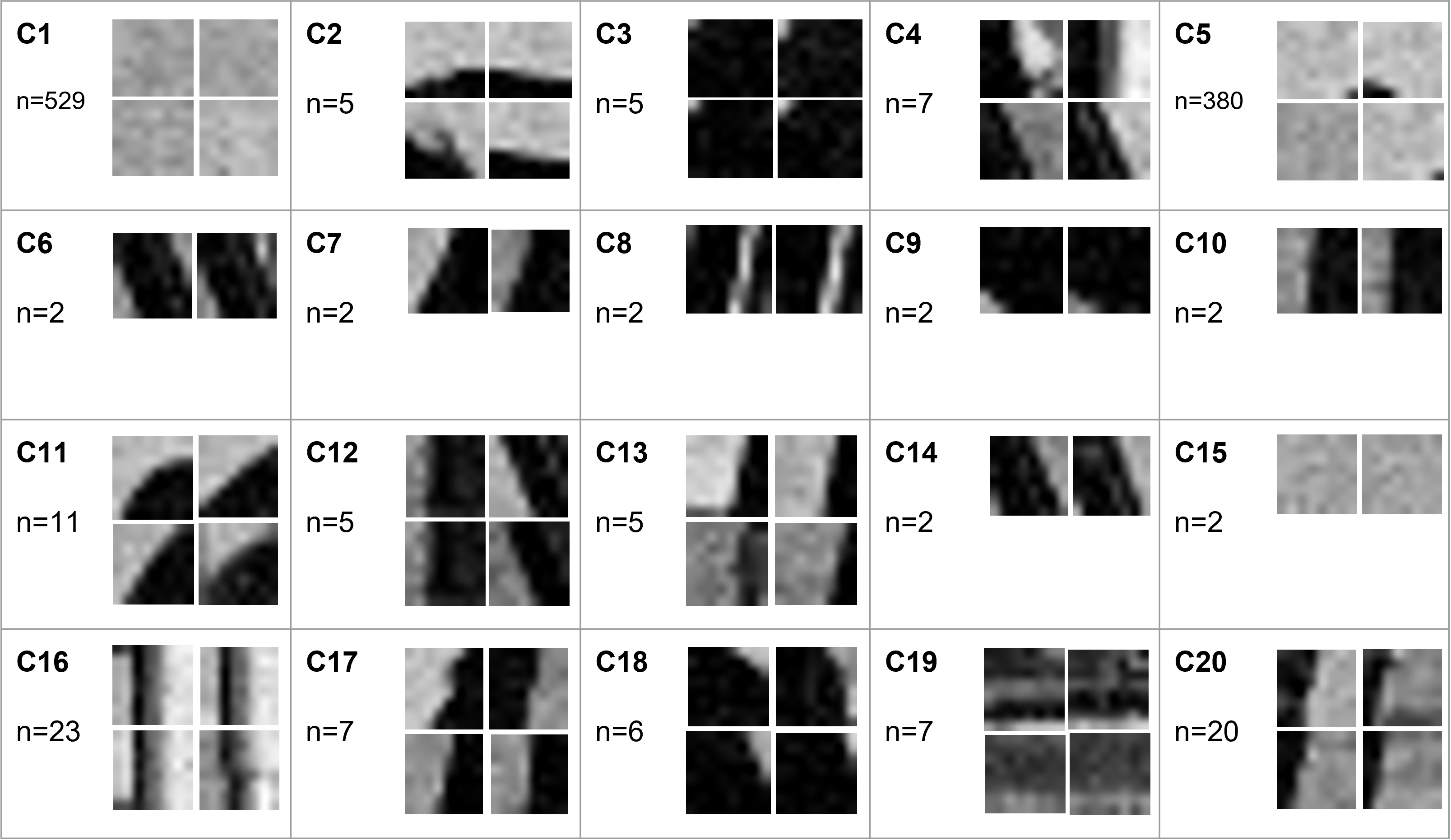}
    
        \includegraphics[width=0.48\textwidth]{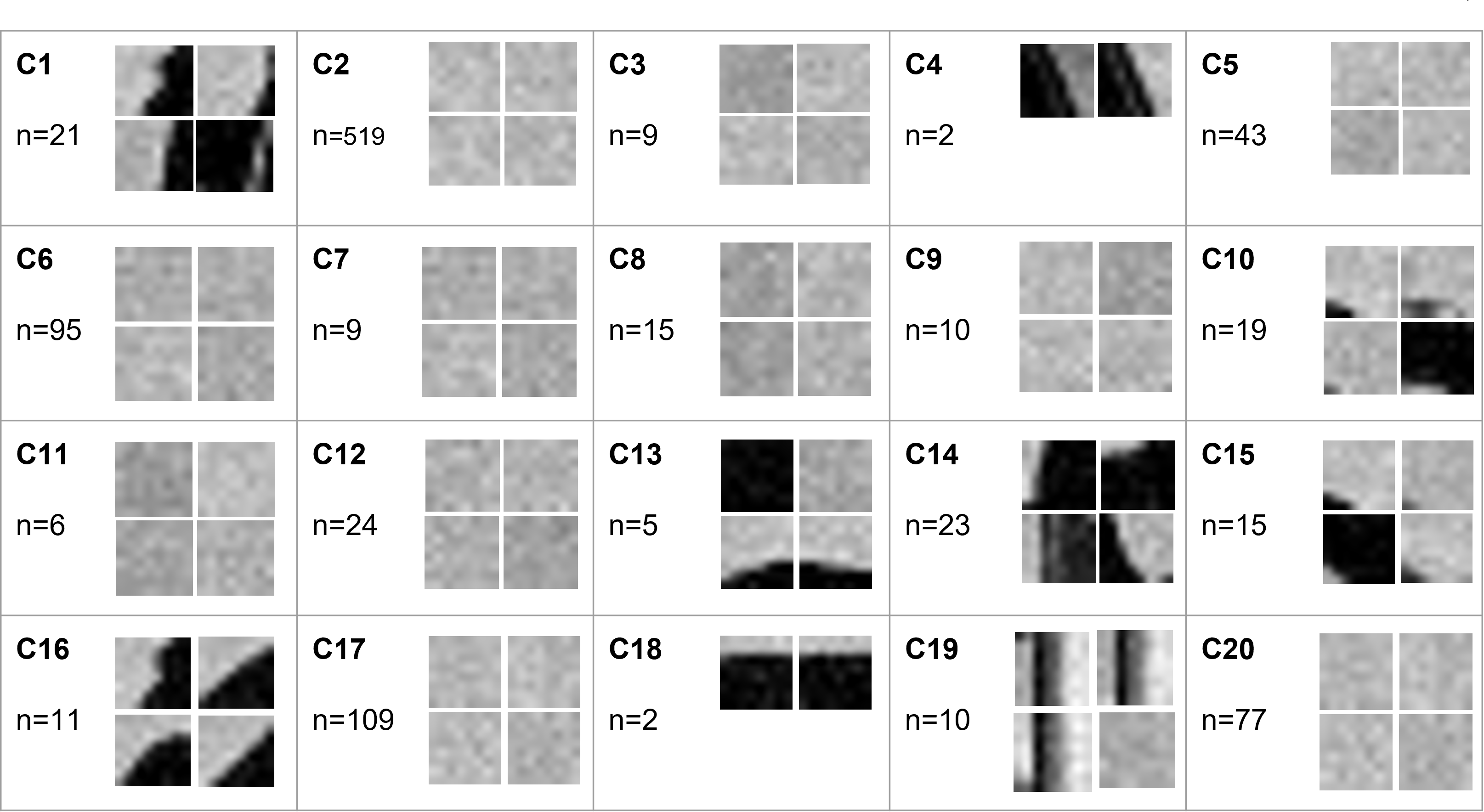}
    
    \caption{Clusters obtained from the {\em Cameraman} image for two competing methods, including four randomly drawn samples from each cluster. For each cluster, $n$ denotes the number of patches in the cluster. {\em Top}: clusters generated using NNC lasso. {\em Bottom}: clusters generated using BPDN. }
    \label{fig:main_figure}
\end{figure}

To provide a qualitative evaluation of the clustering methods, we provide the size $n$ (number of patches) and a randomly chosen set of 4 patches (or fewer if $n < 4$) from each cluster obtained for each method when applied to the {\em Cameraman} image. Figure~\ref{fig:main_figure} shows an example of a comparison between the clustering results obtained when using NNC lasso and BPDN to obtain patch representations. It is clearly observed that the clusters obtained from NNC lasso representations are more distinguishable than those obtained from BPDN representations. In particular, note that NNC lasso's clusters C1 and C5 contain {\em flat} (or nearly flat) patches corresponding to the sky in the {\em Cameraman} image ($n=529$ and 380, respectively), and other clusters have consistent structure in the samples. 

In contrast, the clusters obtained from BPDN representations exhibit flat patches appearing in multiple clusters. Although two of the BPDN clusters, C2 and C17, have similar sizes to those mentioned above ($n=519$ and 109, respectively), there appear to still be about 100 flat patches that are assigned to other clusters -- more specifically, 
a random sampling of cluster contents found that
11 out of 20 clusters returned flat patches. Such an artifact in the clustering distribution is not observed with NNC lasso, and we conjecture that the uniform presence of flat patches among BPDN clusters is due to the positioning of these patches being approximately equidistant to several linear subspaces. Hence, if affine clustering is not explicitly conditioned in the clustering algorithm, the patches are randomly assigned to any of the available clusters.

\begin{figure}[t!]
    \centering
    \includegraphics[width= 0.475\textwidth]{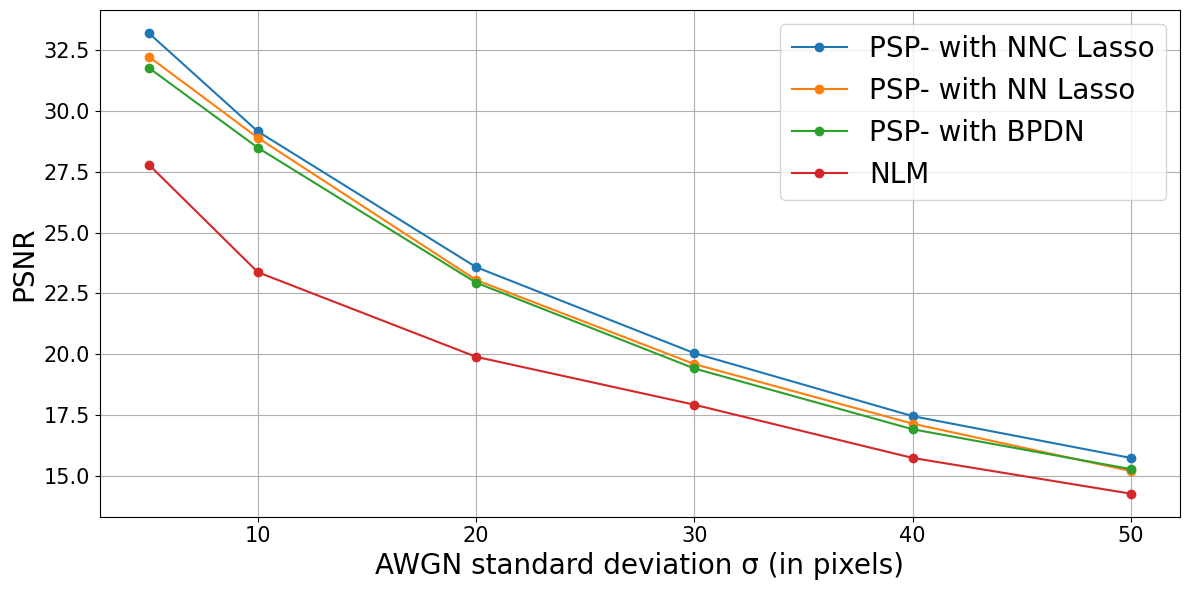}  
    \caption{Denoising comparison of patched affine subspace clustering methods, compared to a baseline NLM algorithm at different noise levels. Evaluations are averaged across all test images.}
    \label{fig:wide_image}
\end{figure}

\begin{figure}[t!]
    \centering
    \includegraphics[width= 0.475\textwidth]{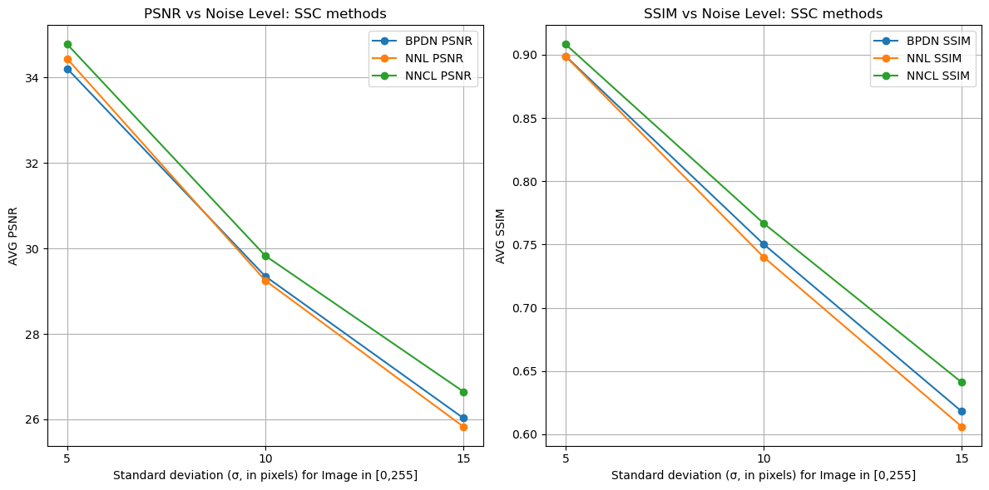}  
    \caption{Comparison among patched subspace clustering methods. Enforcing the affine constraint through NNC lasso consistently improves the denoising performance across noise levels}
    \label{fig:comparing_PSP}
\end{figure}

Next, we summarize the performance of the proposed denoiser algorithms. Figure~\ref{fig:wide_image} shows the improvement in denoising performance gained by patched subspace clustering, 
averaged across all test images. It is
compared to a Non-Local Means (NLM) baseline denoiser implemented in {\em scikit-learn}. 
In our NLM implementation, the filter parameter 
$h$ is set proportionally to the estimated noise level, $h=k\hat{\sigma}$ where the estimate $\hat{\sigma}$ is obtained from the noisy observation. We evaluate several values of $k$ and choose $k=0.01$ based on peak PSNR performance. While PSNR is highest for this configuration, the corresponding reconstructions exhibit significant oversmoothing and appear visually blurred (Figure \ref{fig:cameraman}), highlighting a mismatch between PSNR and perceived image sharpness.
Figure~\ref{fig:comparing_PSP} shows the peak signal-to-noise ratio (PSNR) of the denoised image as a function of the AWGN variance for the different denoising algorithms. The figure shows that the PSP denoiser with NNC lasso consistently outperforms the alternatives; in addition, while PSP with NN lasso outperforms PSP with BPDN for low noise levels, their performances become similar as the noise increases. This behavior correlates with the affineness quality of the different representations -- recall that while NNC lasso provides exact affineness, NC lasso usually provides approximate affineness, and BPDN does not imply affineness in the modeled subspaces. It is also clear that NLM is outperformed by all the PSP approaches.

\section{Conclusion and Future Work}
\label{sec:conclusions}
We considered the use of affine subspaces as a model for image patches and its application to denoising. The affine model is motivated by the non-negativity of the image patches; our numerical results show that affine subspace clustering approaches perform better both in terms of the structure captured by each subspace as well as in terms of the denoising performance. 

The potential of the proposed approaches is significant given that denoisers have found renewed applicability in modern image processing. Examples include image recovery with denoiser-based regularization as well as the so-called plug-and-play approaches to image recovery in inverse problems such as compressive sensing ~\cite{venkatakrishnan2013plug,chan2015adaptive,chan2016plug,reehorst2018regularization,romano2017little,zhang2018nonlocal}.

\vfill
{\bf Acknowledgments:} We thank the IEEE Signal Processing Society's Mentoring Experiences for Underrepresented Young Researchers (ME-UYR) program for their support of this research.

\newpage
\bibliographystyle{IEEEbib}
\bibliography{Alisomar2025}

@inproceedings{venkatakrishnan2013plug,
	author = {Venkatakrishnan, Singanallur V and Bouman, Charles A and Wohlberg, Brendt},
	booktitle = {2013 IEEE Global Conf. Signal and Info. Proc.},
	organization = {IEEE},
	pages = {945--948},
	title = {Plug-and-play priors for model based reconstruction},
	year = {2013}}

@article{reehorst2018regularization,
	author = {Reehorst, Edward T and Schniter, Philip},
	journal = {IEEE Trans. Computational Imaging},
	number = {1},
	pages = {52--67},
	publisher = {IEEE},
	title = {Regularization by denoising: Clarifications and new interpretations},
	volume = {5},
	year = {2018}}

@article{romano2017little,
	author = {Romano, Yaniv and Elad, Michael and Milanfar, Peyman},
	journal = {SIAM J. Imaging Sciences},
	number = {4},
	pages = {1804--1844},
	publisher = {SIAM},
	title = {The little engine that could: Regularization by denoising ({RED})},
	volume = {10},
	year = {2017}}

@article{buades2005review,
	author = {Buades, A. and Coll, B. and Morel, J. M.},
	doi = {10.1137/040616024},
	eprint = {https://doi.org/10.1137/040616024},
	journal = {Multiscale Modeling \& Simulation},
	number = {2},
	pages = {490-530},
	title = {A Review of Image Denoising Algorithms, with a New One},
	url = {https://doi.org/10.1137/040616024},
	volume = {4},
	year = {2005}}

@inproceedings{you2019affine,
	author = {You, Chong and Li, Chun-Guang and Robinson, Daniel P and Vidal, Ren{\'e}},
	booktitle = {IEEE/CVF Int. Conf. Computer Vision ({ICCV})},
	pages = {9915--9924},
	title = {Is an affine constraint needed for affine subspace clustering?},
	year = {2019}}

@inproceedings{wang2022convergence,
	author = {Wang, Peng and Liu, Huikang and So, Anthony Man-Cho and Balzano, Laura},
	booktitle = {Int. Conf. Machine Learning ({ICML})},
	organization = {PMLR},
	pages = {22884--22918},
	title = {Convergence and recovery guarantees of the k-subspaces method for subspace clustering},
	year = {2022}}

@article{soltanolkotabi2014l1,
	author = {Mahdi Soltanolkotabi and Ehsan Elhamifar and Emmanuel J. Cand{\`e}s},
	doi = {10.1214/13-AOS1199},
	journal = {The Annals of Statistics},
	number = {2},
	pages = {669 -- 699},
	publisher = {Institute of Mathematical Statistics},
	title = {{Robust subspace clustering}},
	url = {https://doi.org/10.1214/13-AOS1199},
	volume = {42},
	year = {2014}}

@ARTICLE{dabov20073d,
  author={Dabov, Kostadin and Foi, Alessandro and Katkovnik, Vladimir and Egiazarian, Karen},
  journal={IEEE Transactions on Image Processing}, 
  title={Image Denoising by Sparse {3-D} Transform-Domain Collaborative Filtering}, 
  year={2007},
  volume={16},
  number={8},
  pages={2080-2095},
  doi={10.1109/TIP.2007.901238}}

@INPROCEEDINGS{mairal2009non,
  author={Mairal, Julien and Bach, Francis and Ponce, Jean and Sapiro, Guillermo and Zisserman, Andrew},
  booktitle={2009 IEEE 12th International Conference on Computer Vision}, 
  title={Non-local sparse models for image restoration}, 
  year={2009},
  volume={},
  number={},
  pages={2272-2279},
  doi={10.1109/ICCV.2009.5459452}}

@InProceedings{chen2015external,
author = {Chen, Fei and Zhang, Lei and Yu, Huimin},
title = {External Patch Prior Guided Internal Clustering for Image Denoising},
booktitle = {Proceedings of the IEEE International Conference on Computer Vision (ICCV)},
month = {December},
year = {2015}
}

@ARTICLE{vidal2011subspace,
  author={Vidal, Rene},
  journal={IEEE Signal Processing Magazine}, 
  title={Subspace Clustering}, 
  year={2011},
  volume={28},
  number={2},
  pages={52-68},
  doi={10.1109/MSP.2010.939739}}

@ARTICLE{elhamifar2013sparse,
  author={Elhamifar, Ehsan and Vidal, René},
  journal={IEEE Transactions on Pattern Analysis and Machine Intelligence}, 
  title={Sparse Subspace Clustering: Algorithm, Theory, and Applications}, 
  year={2013},
  volume={35},
  number={11},
  pages={2765-2781},
  doi={10.1109/TPAMI.2013.57}}

@article{gaines2018algorithms,
  title={Algorithms for fitting the constrained lasso},
  author={Gaines, Brian R and Kim, Juhyun and Zhou, Hua},
  journal={Journal of Computational and Graphical Statistics},
  volume={27},
  number={4},
  pages={861--871},
  year={2018},
  publisher={Taylor \& Francis}
}

@INPROCEEDINGS{elhamifar2011sparsity,
  author={Elhamifar, Ehsan and Vidal, René},
  booktitle={2011 49th Annual Allerton Conference on Communication, Control, and Computing (Allerton)}, 
  title={Sparsity in unions of subspaces for classification and clustering of high-dimensional data}, 
  year={2011},
  volume={},
  number={},
  pages={1085-1089},
  doi={10.1109/Allerton.2011.6120288}}

@ARTICLE{li2018on,
  author={Li, Chun-Guang and You, Chong and Vidal, René},
  journal={IEEE Journal of Selected Topics in Signal Processing}, 
  title={On Geometric Analysis of Affine Sparse Subspace Clustering}, 
  year={2018},
  volume={12},
  number={6},
  pages={1520-1533},
  doi={10.1109/JSTSP.2018.2867446}}

@INPROCEEDINGS{ji2014efficient,
  author={Pan Ji and Salzmann, Mathieu and Hongdong Li},
  booktitle={IEEE Winter Conference on Applications of Computer Vision}, 
  title={Efficient dense subspace clustering}, 
  year={2014},
  volume={},
  number={},
  pages={461-468},
  doi={10.1109/WACV.2014.6836065}}

@article{Berg2011SparseOW,
  title={Sparse Optimization with Least-Squares Constraints},
  author={Ewout van den Berg and Michael P. Friedlander},
  journal={SIAM J. Optim.},
  year={2011},
  volume={21},
  pages={1201-1229},
  url={https://pypi.org/project/spgl1/}
}

@article{scikit-learn,
  title={Scikit-learn: Machine Learning in {P}ython},
  author={Pedregosa, F. and Varoquaux, G. and Gramfort, A. and Michel, V.
          and Thirion, B. and Grisel, O. and Blondel, M. and Prettenhofer, P.
          and Weiss, R. and Dubourg, V. and Vanderplas, J. and Passos, A. and
          Cournapeau, D. and Brucher, M. and Perrot, M. and Duchesnay, E.},
  journal={Journal of Machine Learning Research},
  volume={12},
  pages={2825--2830},
  year={2011}
}

@misc{matlabIPT_demoimages,
  title        = {MATLAB and Image Processing Toolbox Demo Images},
  howpublished = {\url{https://www.mathworks.com/products/image-processing.html}},
  note         = {The MathWorks, Inc., Natick, MA, USA},
  year         = {2024}
}

@article{milanfar2012tour,
  title={A tour of modern image filtering: New insights and methods, both practical and theoretical},
  author={Milanfar, Peyman},
  journal={IEEE signal processing magazine},
  volume={30},
  number={1},
  pages={106--128},
  year={2012},
  publisher={IEEE}
}

@article{chatterjee2011patch,
  title={Patch-based near-optimal image denoising},
  author={Chatterjee, Priyam and Milanfar, Peyman},
  journal={IEEE Transactions on Image Processing},
  volume={21},
  number={4},
  pages={1635--1649},
  year={2011},
  publisher={IEEE}
}

@article{elad2006image,
  title={Image denoising via sparse and redundant representations over learned dictionaries},
  author={Elad, Michael and Aharon, Michal},
  journal={IEEE Transactions on Image processing},
  volume={15},
  number={12},
  pages={3736--3745},
  year={2006},
  publisher={IEEE}
}

@inproceedings{zoran2011learning,
  title={From learning models of natural image patches to whole image restoration},
  author={Zoran, Daniel and Weiss, Yair},
  booktitle={2011 international conference on computer vision},
  pages={479--486},
  year={2011},
  organization={IEEE}
}

@inproceedings{gu2014weighted,
  title={Weighted nuclear norm minimization with application to image denoising},
  author={Gu, Shuhang and Zhang, Lei and Zuo, Wangmeng and Feng, Xiangchu},
  booktitle={Proceedings of the IEEE conference on computer vision and pattern recognition},
  pages={2862--2869},
  year={2014}
}

@inproceedings{zhang2018nonlocal,
  title={Nonlocal low-rank tensor factor analysis for image restoration},
  author={Zhang, Xinyuan and Yuan, Xin and Carin, Lawrence},
  booktitle={Proceedings of the IEEE Conference on Computer Vision and Pattern Recognition},
  pages={8232--8241},
  year={2018}
}

@article{portilla2003image,
  title={Image denoising using scale mixtures of Gaussians in the wavelet domain},
  author={Portilla, Javier and Strela, Vasily and Wainwright, Martin J and Simoncelli, Eero P},
  journal={IEEE Transactions on Image processing},
  volume={12},
  number={11},
  pages={1338--1351},
  year={2003},
  publisher={IEEE}
}

@article{chan2016plug,
  title={Plug-and-play ADMM for image restoration: Fixed-point convergence and applications},
  author={Chan, Stanley H and Wang, Xiran and Elgendy, Omar A},
  journal={IEEE Transactions on Computational Imaging},
  volume={3},
  number={1},
  pages={84--98},
  year={2016},
  publisher={IEEE}
}

@inproceedings{chan2015adaptive,
  title={Adaptive patch-based image denoising by EM-adaptation},
  author={Chan, Stanley H and Luo, Enming and Nguyen, Truong Q},
  booktitle={2015 IEEE Global Conference on Signal and Information Processing (GlobalSIP)},
  pages={810--814},
  year={2015},
  organization={IEEE}
}

@article{deledalle2018image,
  title={Image denoising with generalized Gaussian mixture model patch priors},
  author={Deledalle, Charles-Alban and Parameswaran, Shibin and Nguyen, Truong Q},
  journal={SIAM Journal on Imaging Sciences},
  volume={11},
  number={4},
  pages={2568--2609},
  year={2018},
  publisher={SIAM}
}

@article{zhang2010two,
  title={Two-stage image denoising by principal component analysis with local pixel grouping},
  author={Zhang, Lei and Dong, Weisheng and Zhang, David and Shi, Guangming},
  journal={Pattern recognition},
  volume={43},
  number={4},
  pages={1531--1549},
  year={2010},
  publisher={Elsevier}
}

@inproceedings{deledalle2011image,
  title={Image denoising with patch based PCA: local versus global.},
  author={Deledalle, Charles-Alban and Salmon, Joseph and Dalalyan, Arnak S and others},
  booktitle={BMVC},
  volume={81},
  pages={425--455},
  year={2011}
}

@inproceedings{dabov2009bm3d,
  title={BM3D image denoising with shape-adaptive principal component analysis},
  author={Dabov, Kostadin and Foi, Alessandro and Katkovnik, Vladimir and Egiazarian, Karen},
  booktitle={SPARS'09-Signal Processing with Adaptive Sparse Structured Representations},
  year={2009}
}

@article{bradley2000k,
  title={K-plane clustering},
  author={Bradley, Paul S and Mangasarian, Olvi L},
  journal={Journal of Global optimization},
  volume={16},
  number={1},
  pages={23--32},
  year={2000},
  publisher={Springer}
}

@article{lebrun2013nonlocal,
  title={A nonlocal Bayesian image denoising algorithm},
  author={Lebrun, Marc and Buades, Antoni and Morel, Jean-Michel},
  journal={SIAM Journal on Imaging Sciences},
  volume={6},
  number={3},
  pages={1665--1688},
  year={2013},
  publisher={SIAM}
}

@article{wu2013probabilistic,
  title={Probabilistic non-local means},
  author={Wu, Yue and Tracey, Brian and Natarajan, Premkumar and Noonan, Joseph P},
  journal={IEEE Signal Processing Letters},
  volume={20},
  number={8},
  pages={763--766},
  year={2013},
  publisher={IEEE}
}

@article{chan2014monte,
  title={Monte Carlo non-local means: Random sampling for large-scale image filtering},
  author={Chan, Stanley H and Zickler, Todd and Lu, Yue M},
  journal={IEEE transactions on image processing},
  volume={23},
  number={8},
  pages={3711--3725},
  year={2014},
  publisher={IEEE}
}

@inproceedings{roth2005fields,
  title={Fields of experts: A framework for learning image priors},
  author={Roth, Stefan and Black, Michael J},
  booktitle={2005 IEEE Computer Society Conference on Computer Vision and Pattern Recognition (CVPR'05)},
  volume={2},
  pages={860--867},
  year={2005},
  organization={IEEE}
}

@article{van2014student,
  title={The student-t mixture as a natural image patch prior with application to image compression.},
  author={van den Oord, A{\"a}ron and Schrauwen, Benjamin},
  journal={J. Mach. Learn. Res.},
  volume={15},
  number={1},
  pages={2061--2086},
  year={2014}
}

@inproceedings{liu2010robust,
  title={Robust subspace segmentation by low-rank representation},
  author={Liu, Guangcan and Lin, Zhouchen and Yu, Yong},
  booktitle={Proceedings of the 27th international conference on machine learning (ICML-10)},
  pages={663--670},
  year={2010}
}

@inproceedings{li2015structured,
  title={Structured sparse subspace clustering: A unified optimization framework},
  author={Li, Chun-Guang and Vidal, Rene},
  booktitle={Proceedings of the IEEE conference on computer vision and pattern recognition},
  pages={277--286},
  year={2015}
}

\end{document}